# Evolution of Critical Current Density in CaKFe$_4$As$_4$ with La-doping

Yangsong Chen[1,*], Chunlei Wang[1,2], Yuhang Zu[1], Yuto Kobayashi[1], Ataru Ichinose[3], Ryosuke Sakagami[1], and Tsuyoshi Tamegai[1]

[1]Department of Applied Physics, The University of Tokyo, 7-3-1 Hongo, Bunkyo-ku, Tokyo 113-8655, Japan

[2]Department of Physics and Electronic Engineering, Xinyang Normal University, Xinyang 464000, China

[3]Grid Innovation Research Laboratory. Central Research Institute of Electric Power Industry, 2-6-1 Nagasaka, Yokosuka, Kanagawa 240-0196, Japan

E-mail: C.yangsong@outlook.com

**Abstract**

Single crystals of (Ca$_{1-x}$La$_x$)KFe$_4$As$_4$ ($0 \leq x \leq 0.16$) have been grown by using the self-flux method, and the evolution of physical properties including the critical current density ($J_c$) with La-doping has been investigated. $T_c$ decreases monotonically with increasing $x$, while $J_c$ at the same temperature and magnetic field increases initially and reach its maximum at $x = 0.082$. The increase in $J_c$ is more obvious at low temperatures and high fields. At $T$ = 5 K and $H$ = 40 kOe, $J_c$ reaches 0.34 MA/cm$^2$, which is ~4 times larger than that for pure crystals. It is also found that anomalous temperature dependence of $J_c$ in CaKFe$_4$As$_4$ is wiped away as the La content is increased. However, $J_c$ shows non-monotonic field dependence (peak effect) at high fields in crystals with large $x$. In addition, we found that despite weak anisotropy of $H_{c2}$, there is extremely large anisotropy of $J_c$ up to ~15, which is most likely caused by novel planar defects in the crystal, similar to CaKFe$_4$As$_4$. $J_c$ characteristics in (Ca$_{1-x}$La$_x$)KFe$_4$As$_4$ with disorder outside FeAs planes is compared with that in CaK(Fe$_{1-x}$Co$_x$)$_4$As$_4$ with disorder within FeAs planes.

## 1. Introduction

Since the discovery of LaFeAs(O,F) with $T_c$ ~ 26 K, iron-based superconductors (IBSs) have been extensively studied as a new class of high-temperature superconductors[1]. The 122-type of IBSs (($AE,A$)Fe$_2$As$_2$, $AE$: Sr and Ba, $A$: Na, K, Rb)) have been investigated as promising materials for applications because of their larger critical current density ($J_c$) exceeding $1 \times 10^6$ A/cm$^2$ at 4.2 K, large upper critical field ($H_{c2}$) larger than 700 kOe[2,3], and smaller anisotropy ($\gamma \sim 1-2$)[4]. Recently, another promising IBSs (1144-type), CaKFe$_4$As$_4$ as a representative material, have been discovered[5], and attract much interest due to their characteristic features such as stoichiometric superconductivity and anomalous behavior of $J_c$ in the presence of novel planar defects[6]. A novel magnetic state with hedgehog spin structure[7,8] has also been studied with its effect on superconducting gap [9] together with its pressure dependence[10]. Compared with 122-type compounds, Ca and K layers in CaKFe$_4$As$_4$ stack alternatively along the $c$-axis[5,11]. Superconductivity shows up at ~35.5 K similar to slightly over-doped Ba$_{0.5}$K$_{0.5}$Fe$_2$As$_2$[12–14] and large $H_{c2}$ similar to (Ba,K)Fe$_2$As$_2$ is reported [13]. Despite alternate stacking of Ca and K along the $c$-axis, CaKFe$_4$As$_4$ shares various common properties with the 122-type compounds, such as large $J_c$. However, in CaKFe$_4$As$_4$ at high magnetic fields, $J_c$ presents a very different temperature dependence. At high magnetic fields, the $J_c$ in CaKFe$_4$As$_4$ shows nonmonotonic temperature dependence, namely $J_c$ at high temperatures is larger than that at low temperatures[14–16]. This novel feature suggests a possibility that CaKFe$_4$As$_4$ can be utilized in a broad temperature range.

The $J_c$ is determined by pinning of vortices in superconductors. In addition to the pinning naturally present in the crystal, pinning centers can be artificially engineered into superconductors through defects generated by high-energy particle irradiation[17–20] or chemical doping[21,22]. In previous studies on Ni-doped CaK(Fe$_{1-x}$Ni$_x$)$_4$As$_4$[23] and Co-doped CaK(Fe$_{1-x}$Co$_x$)$_4$As$_4$[16], $J_c$ was enhanced appreciably by increasing the doping level in spite of the suppression of $T_c$ due to disorder in the crystal. In reference to (Ba$_{1-x}$K$_x$)Fe$_2$As$_2$ with the highest $T_c$ (~ 38.5 K) for $x$ = 0.4, CaKFe$_4$As$_4$ can be considered to be over-doped, equivalent to $x$ = 0.5. According to the phase diagram of (Ba$_{1-x}$K$_x$)Fe$_2$As$_2$ [24], introduction of electrons in CaKFe$_4$As$_4$ can be a way to enhance $T_c$. However, in recent studies, substitution of alkali metal or alkaline earth metal sites leads to changes in $T_c$, which may come from changes in the lattice constant [25–27]. In these works, some substituted polycrystalline samples have been grown, and no enhance of $T_c$ was reported. It has been reported that $J_c$ in CaKFe$_4$As$_4$ shows non-monotonic temperature and magnetic field dependences. In addition, $J_c$ for different field orientation is extremely anisotropic[6,28]. High-resolution electron microscope observations revealed the presence of planar defects perpendicular to the $c$-axis, consisting mainly of KFe$_2$As$_2$ intergrowth [29]. The role of these planar defects to the anomalous behavior of $J_c$ is under intensive debate.

In this work, we have successfully grown high-quality single crystals of La-doped (Ca,La)KFe$_4$As$_4$ with various doping levels, keeping sharp superconducting transition with flat magnetization at low temperatures even after La-doping. We evaluated superconducting properties for La-doped (Ca,La)KFe$_4$As$_4$ including the effect of chemically introduced point defects on $T_c$ and $J_c$, which could be related to the novel planar defects[6,28].

## 2. Experiments

Single crystals of La-doped (Ca$_{1-x}$La$_x$)KFe$_4$As$_4$ with $x$ up to 0.16 were grown by the self-flux method with FeAs flux. Ca granules (99.5%), La granules (99.5%), K ingots (99.5%), and FeAs powder were used as starting materials. FeAs was prepared by sealing stoichiometric amounts of As grains (7N) and Fe powder (99.9%) in an evacuated quartz tube. It was heated up to 500 °C for 10 h and then heated up to 700 °C for 40 h. A mixture with a ratio of Ca : La : K : FeAs = (1-$x$): $x$ : 1.1: 10 was placed in an alumina crucible in an argon-filled glove box and nominal $x$ = 0, 0.03, 0.075, 0.1, and 0.2. The alumina crucible was then sealed in a stainless-steel tube by mechanically swaging the cap. For $x \leq 0.1$, the stainless-steel tube was sealed in an evacuated quartz tube to avoid oxidation. The whole assembly was heated up to 650 °C for 5 h, and then heated up to 1160 °C for 5 h. It was cooled down to 1050 °C in 5 h and slowly cooled down to 930 °C at a rate of 1.5 °C/h. For $x > 0.1$, the stainless-steel tube was placed in nitrogen flow during the growth. Different from the previous report[30,31], the highest temperature was increased up to 1180 °C and slowly cooled down to 910 °C for the crystal growth.

After the growth, several samples were examined by a scanning electron microscope (S-4300, Hitachi High Technologies), and elemental analyze were conducted using energy-dispersive X-ray spectroscopy (EDX) with EMAX X-act (HORIBA). The experiment was carried out at room temperature with an electron accelerating voltage of 20 kV. The magnetization of single crystals was measured for the characterization of $T_c$ and magnetic $J_c$ by a superconducting quantum interference device magnetometer (SQUID) (MPMS-5XL, Quantum Design). $T_c$ was estimated from magnetization measurements for field parallel to the $c$-axis. $J_c$ was evaluated from the results of magnetization measurements using the extended Bean model[32],

$$J_c = \frac{20\Delta M}{a\left(1 - \frac{a}{3b}\right)} \quad (a < b), \tag{1}$$

where $\Delta M$ (emu cm$^{-3}$) is the difference of magnetization when the applied field is swept down and up. $a$ (cm) and $b$ (cm) are width and length of the single crystal. For the evaluation of two components of $J_c$ for $H//ab$-plane, samples were carefully cut into a rectangular shape using a wire saw. Electrical resistivity along the $ab$-plane ($\rho_{ab}$) and Hall coefficient ($R_H$) were measured by the four-probe method under a magnetic field parallel to the $c$-axis, and the resistance was measured by using an AC resistance bridge (LR-700, Linear Research). Low temperature conditions and static magnetic fields were provided by MPMS.

## 3. Results

It has been reported that actual compositions of IBS single crystals grown using the flux method can be different from the nominal composition, as evidenced by the case of Ba(Fe,Co)$_2$As$_2$[33]. We examined the actual composition of our crystals using SEM-EDX for both La (La/(La+Ca) as actual $x$) and Co dopings (Co/(Co+Fe) as actual $x$). We used the same Co-doped crystals as those used in the previous study [16]. It turns out that the actual doping levels are lower than the nominal ones. Table 1 summarizes the comparison of the nominal and actual doping levels of our crystals. In the rest of the paper, we use the actual $x$ for both La and Co doping levels.

| | | | | | | |
|---|---|---|---|---|---|---|
| La-doping | Nominal $x$ | 0.05 | 0.075 | 0.1 | 0.2 | |
| | Actual $x$ | 0.026 (±0.004) | 0.082 (±0.005) | 0.097 (±0.002) | 0.164 (±0.003) | |
| Co-doping | Nominal $x$ | 0.01 | 0.03 | 0.05 | 0.07 | 0.09 |
| | Actual $x$ | 0.008 (±0.004) | 0.023 (±0.004) | 0.036 (±0.002) | 0.046 (±0.005) | 0.073 (±0.002) |

Table 1. Comparison of the nominal and actual doping levels for (Ca$_{1-x}$La$_x$)KFe$_4$As$_4$ and CaK(Fe$_{1-x}$Co$_x$)$_4$As$_4$ crystals.

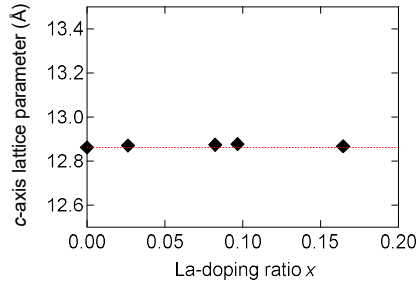

FIG. 1. $c$-axis lattice parameter as a function of La-doping levels for (Ca,La)KFe$_4$As$_4$.

Figure 1 shows the change of $c$-axis lattice parameter for single crystals after doping. We found that the $c$-axis lattice parameter changes very little with the doping. Due to very similar the ionic radii of La and Ca, La-doping does not change the lattice parameter significantly. A similar very weak change of $c$-axis lattice parameter upon La-doping is also reported in the work on (Ca,La)Fe$_2$As$_2$.[34].

Figure 2(a) shows the temperature dependence of normalized magnetization ($M$) at 5 Oe for (Ca$_{1-x}$La$_x$)KFe$_4$As$_4$ single crystals with different La-doping levels ($x$ = 0, 0.026, 0.082, 0.097, and 0.164). As shown in Fig. 2(a), $T_c$ gradually decreases, while $\Delta T_c$, which is defined as the difference between the starting point (90% of the normal state value) and the end point (10% of the normal state value) of the superconducting transition, gradually increases with increasing $x$. $\Delta T_c$ is less than 1 K for $x \leq 0.082$, while it increases to 1.5 K and 3 K for $x$ = 0.097 and 0.164, respectively. The main cause of the broadening of $\Delta T_c$ is attributed to the increasing inhomogeneity of doping, which becomes more significant with increasing the La-doping level. The superconducting transition is also confirmed by the temperature dependence of electrical resistivity as shown in Fig. 2(b). We can difine the residual resistivity ratio (RRR) by using the resistivity values just above $T_c$ and that at room temperature, $\rho(300\ K)/\rho(T_c$ onset). For CaKFe$_4$As$_4$, RRR is ~15, consistent with former reports [6, 7]. After La-doping, RRR is reduced to ~8 at $x = 0.164$. The suppression of RRR after the La-doping is caused by the increase in resistivity at low temperatures due to increased scattering rate. Figure 2(c) shows the temperature dependence of normalized magnetization ($M$) at 5 Oe for CaK(Fe$_{1-x}$Co$_x$)$_4$As$_4$ single crystals with different Co-doping levels ($x = 0, 0.008, 0.023, 0.036, 0.046$, and 0.073). As shown in Fig. 2(c), $T_c$ decreases rapidly with increasing $x$.

Figure 2(d) shows the temperature dependence of the Hall coefficient ($R_H$) for (Ca$_{1-x}$La$_x$)KFe$_4$As$_4$ ($x = 0.082$). Since CaKFe$_4$As$_4$ possesses undoubtedly a multiband electronic structure and is dominated by hole carriers[11], upon introduction of electrons via La-doping, the absolute value of the Hall coefficient becomes larger than that in CaKFe$_4$As$_4$, which corresponds to $x = 0.5$ in (Ba$_{1-x}$K$_x$)Fe$_2$As$_2$. The Hall coefficient shows clear temperature dependence and the value at 40 K is more than twice compared with than that at 300 K. Both (Ca$_{1-x}$La$_x$)KFe$_4$As$_4$ and CaK(Fe$_{1-x}$Co$_x$)$_4$As$_4$ correspond to electron doping into CaKFe$_4$As$_4$. The difference between the two cases is the location of doped elements. Namely, it is outside the FeAs planes in the former case, while it is within the FeAs planes.

Figure 2(e) shows the $T_c$ as functions of doping level $x$ for (Ca$_{1-x}$La$_x$)KFe$_4$As$_4$ and CaK(Fe$_{1-x}$Co$_x$)$_4$As$_4$ single crystals [15]. As the doping level increases, $T_c$ is obviously suppressed in both cases. Since the amount of electrons introduced with the same doping level is not the same, it is more reasonable to use the valence of iron as the $x$-axis. In Fig. 2(f), $T_c$ as functions of the valence of iron for (Ca$_{1-x}$La$_x$)KFe$_4$As$_4$ and CaK(Fe$_{1-x}$Co$_x$)$_4$As$_4$ single crystals are compared. Compared with the Co-doping, the suppression of $T_c$ is weaker in the case of La-doping. It implies that electron doping outside FeAs planes introduces much less disorder in the system. However, since we bring the system towards the optimum doping with the La doping, we may expect the increase of $T_c$. The fact that we did not observe such an increase in $T_c$ indicates that the disorder introduced by the La doping is stronger to compensate the increase in $T_c$ towards optimum doping.

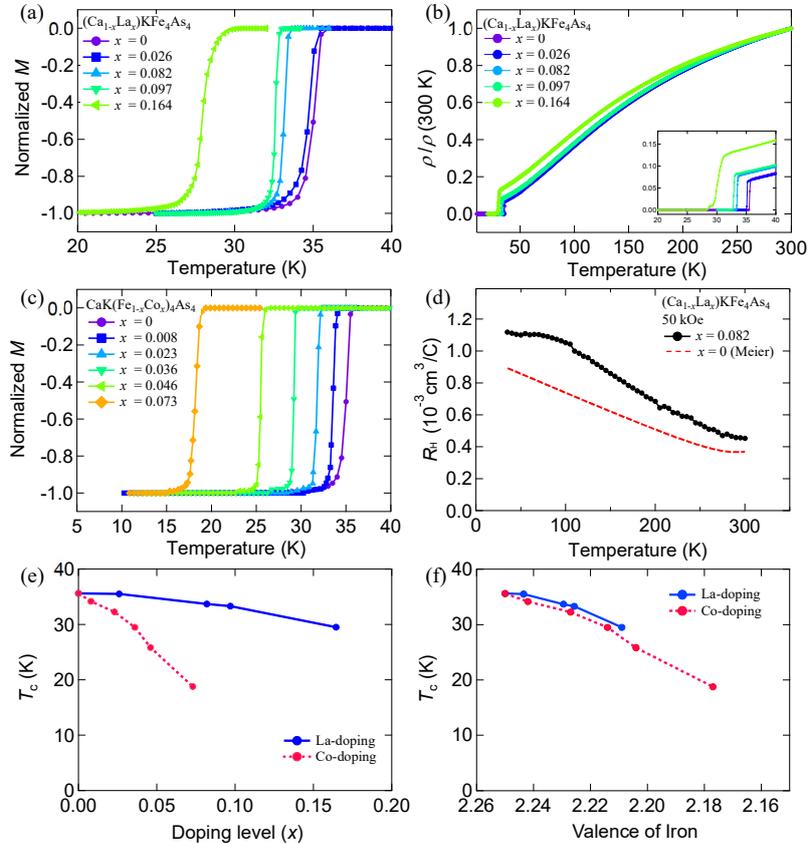

FIG. 2. Temperature dependence of (a) normalized magnetization ($M$) at 5 Oe and (b) normalized resistivity (($\rho/\rho(300 K)$)) for (Ca$_{1-x}$La$_x$)KFe$_4$As$_4$ single crystals with different La-doping levels ($x = 0, 0.026, 0.082, 0.097$, and 0.164). The inset shows

a blow-up of $\rho/\rho(300\text{ K}) - T_c$ near $T_c$. (c) Temperature dependence of normalized magnetization ($M$) at 5 Oe for CaK(Fe$_{1-x}$Co$_x$)$_4$As$_4$ single crystals with different Co-doping levels ($x$ = 0, 0.008, 0.023, 0.036, 0.046, and 0.073). (d) Temperature dependence of Hall coefficient $R_H$ for (Ca$_{1-x}$La$_x$)KFe$_4$As$_4$ ($x$ = 0.082). Data for CaKFe$_4$As$_4$ taken from ref.[11] is also plotted. (e) $T_c$ as functions of doping level $x$ for (Ca$_{1-x}$La$_x$)KFe$_4$As$_4$ and CaK(Fe$_{1-x}$Co$_x$)$_4$As$_4$ single crystals. (f) $T_c$ as functions of valence of iron for (Ca$_{1-x}$La$_x$)KFe$_4$As$_4$ and CaK(Fe$_{1-x}$Co$_x$)$_4$As$_4$ single crystals.

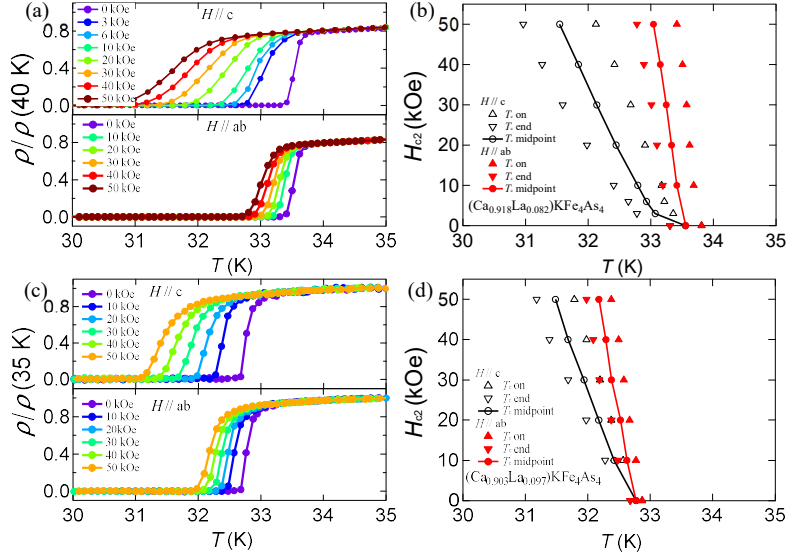

FIG. 3. Temperature dependence of normalized electrical resistivity in (Ca$_{1-x}$La$_x$)KFe$_4$As$_4$ single crystals with (a) $x$ = 0.082 and (c) $x$ = 0.097 under various magnetic fields parallel to the $c$-axis and $ab$-plane. (b) $H_{c2}$ evaluated from temperature-dependent resistivity presented in (a). (d) $H_{c2}$ evaluated from temperature-dependent resistivity presented in (c).

Next, we focus on the properties of (Ca$_{1-x}$La$_x$)KFe$_4$As$_4$ single crystals with $x$ = 0.082, which has the largest $J_c$ in this series as we discuss later. Figs. 3 (a) and (c) shows the in-plane resistivity ($\rho$) for $H//c$-axis and $H//ab$-plane measured at various magnetic fields up to 50 kOe. The onset, midpoint, and endpoint of the superconducting transition were determined by 90%, 50%, and 10% of the normal state value, respectively. $H_{c2}$ is defined by the midpoint of the resistive transition. The $H_{c2}$ at $T = 0$ K is evaluated using WHH formula, leading to $H_{c2}(0) \sim 740$ kOe for $x$ = 0.082 and $H_{c2}(0) \sim 900$ kOe for $x$ = 0.097. We find that in Fig.3 (a), for sample for $x$ = 0.082, the change in $T_c$ is very large when the magnetic field goes from 0 kOe to 10 kOe, which may lead to an overestimation of $H_{c2}(0)$. Compared with the value for CaKFe$_4$As$_4$ $H_{c2}(0) = 900$ kOe [11], and based on the data of sample for $x$ = 0.097, we believe that doping will not significantly affect $H_{c2}(0)$. According to Figs. 3 (b) and (d), we can clearly see that the rate of change of transition for $H//c$-axis is much greater than that for $H//ab$-plane. By fitting the linear part for the temperature dependence of $H_{c2}(T)$, we estimated $\gamma$ is ~3.12 for sample with $x$ = 0.082 at 32 K (~1.5 K close to $T_c$) and ~2.21 for sample for $x$ = 0.097 at 31 K (~1.5 K close to $T_c$). As we mentioned before, for sample for $x$ = 0.082, $H_{c2}(T)$ has a certain offset when it is close to $T_c$. We believe that this offset is the reason why we get a larger number when estimating the $\gamma$ value. Compared to the previously reported value of $\gamma \sim 2.3$ at 34 K in CaKFe$_4$As$_4$ ($\gamma$ is re-evaluated using the 50% criterion)[6,11], $\gamma$ does not change more after doping. It makes a good contrast with $\gamma \sim 2$-$3$ in (Ba$_{1-x}$K$_x$)Fe$_2$As$_2$ near $T_c$, where $\gamma$ is nearly independent of the doping level [31]. Therefore, regardless of the data of the $x$ = 0.082 sample, we believe that doping does not substantially change the anisotropy.

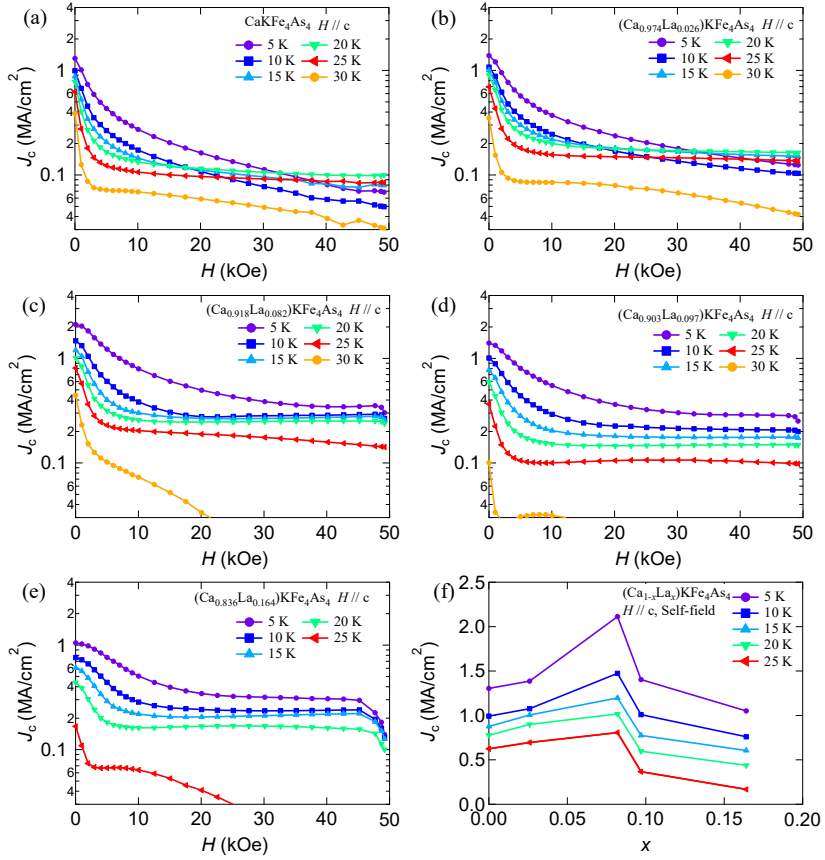

FIG. 4. Magnetic field dependences of $J_c$ at different temperatures in $(Ca_{1-x}La_x)KFe_4As_4$ for different La-doping levels (a) $x = 0$, (b) $x = 0.026$, (c) $x = 0.082$, (d) $x = 0.097$, and (e) $x = 0.164$. (f) Doping level dependences of $J_c$ under the self-field in $(Ca_{1-x}La_x)KFe_4As_4$ at different temperatures.

In-plane $J_c$ in $(Ca_{1-x}La_x)KFe_4As_4$ single crystals with different La-doping levels ($x = 0, 0.026, 0.082, 0.097,$ and $0.164$) was evaluated from magnetization measurements by applying the magnetic field along the $c$-axis. Figures 4(a)-(e) show the magnetic field dependences of $J_c$ at various temperatures in $(Ca_{1-x}La_x)KFe_4As_4$ with $x = 0, 0.026, 0.082, 0.097,$ and $0.164$, respectively. In samples with $x = 0$ and $x = 0.026$, $J_c$ shows novel non-monotonic temperature dependence at high fields. Namely, $J_c - H$ curves at low temperatures crosses those at higher temperatures. Such crossings of $J_c - H$ curves have also been reported in previous studies on $CaKFe_4As_4$ [7,9,14]. Meanwhile, crossing points in Fig. 4(b) are shifted to lower fields compared with those in Fig. 4(a). This shows that small amount of La-doping improves $J_c$ at high temperatures more significantly. It is remarkable that $J_c$ at high temperatures (15 K, 20 K, and 25 K) shows very weak magnetic field dependence at high fields. For $x \geq 0.082$, $J_c$ does not decrease so much at low temperatures, and the non-monotonic temperature dependence of $J_c$ disappears as shown in Figs. 3 (c), (d), and (e). We also note that as the doping level gradually increases, $J_c$ at low fields ($H \sim 10$ kOe) and high fields ($H > 40$ kOe) gradually approach to each other, especially at low temperatures. This change also indicates that doping is more effective in improving $J_c$ under high magnetic fields at low temperatures than at higher temperatures. Figure 4(f) shows the doping level dependence of $J_c$ under the self-field in $(Ca_{1-x}La_x)KFe_4As_4$. It is clear that $J_c$ is sharply peaked at $x \sim 0.08$.

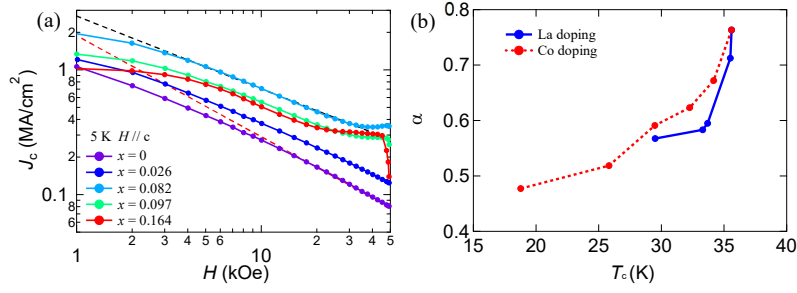

FIG. 5. (a) Magnetic field dependence of magnetic $J_c$ in double logarithmic scale in $(Ca_{1-x}La_x)KFe_4As_4$ single crystals with different La-doping level ($x = 0$, 0.026, 0.082, 0.097, and 0.164 for the field parallel to the $c$-axis at 5 K. Red and black dashed lines show examples of linear fitting to the data for $x = 0$ and $x = 0.082$, respectively. (b) α at 5 K as functions of $T_c$ in $(Ca_{1-x}La_x)KFe_4As_4$ and $CaK(Fe_{1-x}Co_x)_4As_4$ single crystals.

Figure 5(a) shows $J_c$ at $T = 5$ K as functions of magnetic field in double-logarithmic scale for different samples. As the La-doping level is increased, $J_c$ initially increases, reaching its maximum at $x = 0.082$ and then decreases. At the same time, it can be seen that the La-doping significantly improves the performance of $J_c$ at high fields and $J_c$ does not decreases so much at $x \geq 0.082$. Compared with $x = 0$, $J_c$ at $T = 5$ K in the sample with $x = 0.082$ is ~5 times larger at 40 kOe. As the doping level continues to increase, $J_c$ does not continue to increase but decreases. $J_c$ is determined by two factors: one is the strength of the superconductivity. The other is the number of defects in the sample. Samples with stronger superconductivity generally have stronger $J_c$, and more defects enhance the pinning effect, thereby enhance $J_c$. However, when defects are introduced, their superconductivity is usually suppressed. The maximum value of $J_c$ often requires a balance between these two factors. We consider that the gradual suppression of $J_c$ for $x \geq 0.082$ is caused by the suppression of superconductivity with larger doping levels. For some superconductors, a power-law dependence between $J_c$ and $H$, $J_c \propto H^{-\alpha}$, at high fields is reported. For most of pristine IBSs, the power-law exponent close to 5/9, which is interpreted as a result of the presence of sparse strong pinning centers, is confirmed [35]. We find linear part in Fig. 5(a), and can extract. α ~ 0.76 for $CaKFe_4As_4$ at 5 K, which is much larger than α = 5/9. As the La-doping level increases, the value of α continuously decreases, indicating the strengthening of pinning. Figure 5(b) shows α at 5 K as functions of $T_c$ in $(Ca_{1-x}La_x)KFe_4As_4$ and $CaK(Fe_{1-x}Co_x)_4As_4$ single crystals. Compared with Co-doping in $CaK(Fe_{1-x}Co_x)_4As_4$, decline of α in La-doped sample is faster at similar $T_c$, indicating that La-doping enhances pinning more efficiently than Co-doping.

The irreversible magnetization for magnetic field parallel to the $ab$-plane is controlled by two kinds of critical current flowing in-plane ($J_{c2}$) and out-of-plane ($J_{c3}$) as shown in the inset of Fig. 6(c) and (d)[6]. These two $J_c$ components can be evaluated by magnetic measurements with fields applied along two orthogonal directions for a thin rectangular superconductor. The magnetic field dependences of in-plane magnetization at various temperatures for fields along the short ($M_1$) and long ($M_1$) edges of the sample are shown in Figs. 5(a) and (b), respectively. A notable feature of these hysteresis loops is the presence of sharp dips close to $H \sim 0$ in both cases, which is similar to the case of $CaKFe_4As_4$[6]. These features are, in a sense, similar to the case of IBSs with columnar defects introduced by heavy-ion irradiation[19]. However, unlike the case of columnar defects, the characteristic field for the dip is not proportional to $J_c$, suggesting that it is controlled by some factors other than the self-field. In $CaKFe_4As_4$, the presence of novel planar defects parallel to the $ab$-plane are confirmed, and the average separation of these planar defects is consistent with the characteristic field for the dip[6]. Figures 6(c) and (d) shows the magnetic field dependence of $J_{c2}$ and $J_{c3}$, respectively. Obviously, $J_{c2}$ is extremely large, reaching 9.45 MA/cm$^2$ at 5 K and 10 kOe. On the other hand, $J_{c3}$ is about 8 times smaller than $J_{c2}$. For an example, at 5 K and 10 kOe, $J_c$, $J_{c2}$, and $J_{c3}$ are 1.22, 9.45 and 0.64 MA/cm$^2$, respectively.

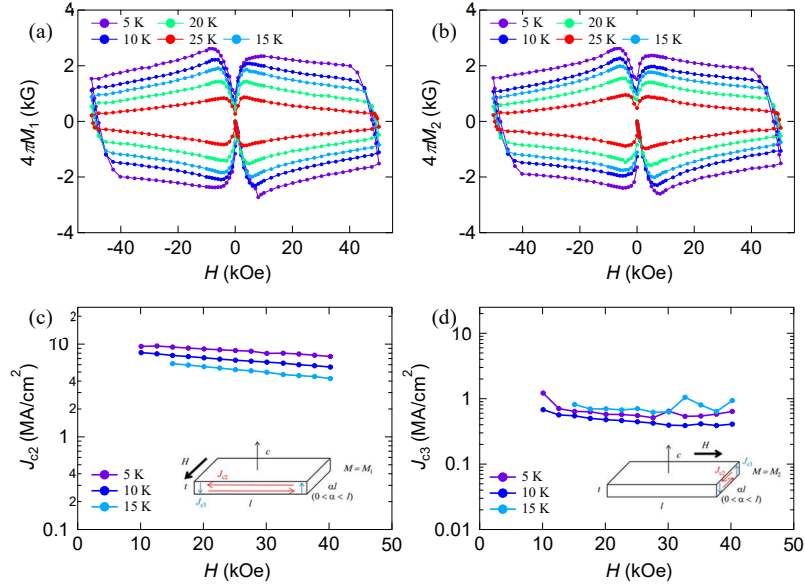

FIG. 6. Magnetic field dependences of in-plane magnetization for fields parallel to the *ab*-plane along (a) short ($M_1$) and (b) long ($M_2$) edges of (Ca$_{1-x}$La$_x$)KFe$_4$As$_4$ single crystals with $x = 0.082$. Magnetic field dependence of (c) $J_{c2}$ and (d) $J_{c3}$ at different temperatures.

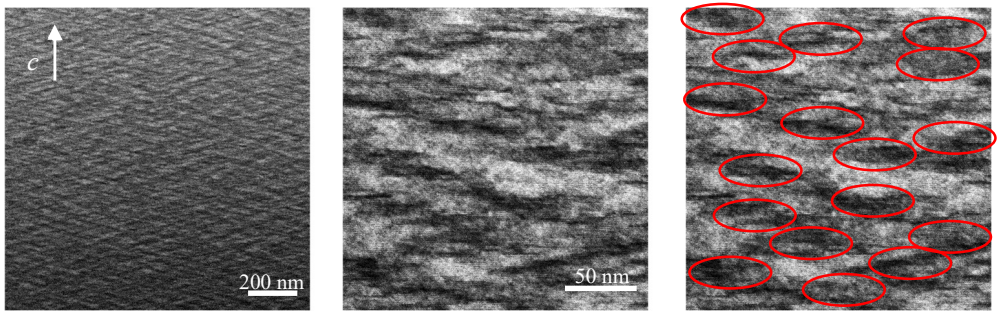

FIG. 7. (a) STEM image of (Ca$_{1-x}$La$_x$)KFe$_4$As$_4$ ($x = 0.082$) for an electron beam injected along the *ab*-plane. (b) A similar STEM image to (a) with higher magnification. (c) The same image as in (b) showing the defect clusters.

Presence of dense planar defects similar to those found in CaKFe$_4$As$_4$ is confirmed by a high-resolution scanning transmission electron microscope (STEM) observation as shown in Fig. 7(a). Higher-resolution image shown in Fig. 7(b) indicates that shapes of individual defects are more irregular compared with those in CaKFe$_4$As$_4$. Several planar defects form a cluster of $\sim 50 \times 20$ nm$^2$, and those clusters are arranged in disordered manner. In CaKFe$_4$As$_4$, there are two different kinds of planar defects that are considered, KFe$_2$As$_2$ [6,29] and CaFe$_2$As$_2$ [28]. Compared with the CaFe$_2$As$_2$ planar defect found in previous studies, the plane defect in our image extends in two different directions and is not parallel to the ab plane. Therefore, we tend to believe that the plane defect in the sample is KFe$_2$As$_2$. In pure crystal, each defect is $\sim 50 \times 10$ nm$^2$, and is separated by $\sim 50$ nm and $\sim 40$ nm along the *ab*-plane and *c*-axis, respectively [6]. But in doped crystal, several planar defects form a cluster of $\sim 50 \times 20$ nm$^2$. We found that the planar defect clusters in the doped samples were larger than those in the pure samples. We estimated their size by eye and found that the planar defect clusters in the doped samples seemed to be twice as large. This means that doping also caused more crystal dislocations and introduced more defects during growth, which may also be the reason for the suppression of $T_c$. If we identify individual defects cluster as shown in Fig. 7(c), the average density of these clusters of planar defects is ~425 $\mu m^{-2}$, which corresponds to the characteristic field of 12 kOe. Given the ambiguity of the density of defect clusters, this characteristic field agrees reasonably well with the characteristic field for the dip of ~7 kOe, which supports our interpretation of dip as originated from the matching effect of vortices to the planar defects. Since these defect clusters are elongated along the $ab$-plane, they will hinder vortex motion along the $c$-axis more effectively, helping to make $J_{c2}$ larger. The same geometry of defect clusters blocks the supercurrent flow along the $c$-axis more effectively, making $J_{c3}$ smaller.

## 4. Discussion

In this study, although electrons were successfully introduced in $CaKFe_4As_4$ by La doping, $T_c$ did not increase at small doping levels. At larger La doping levels, the carrier density exceeds the optimal level of 0.2 holes per Fe, and $T_c$ is suppressed as we anticipated. Although the ionic radius of $Ca^{+2}$ (100 pm) is very similar to that of $La^{+3}$ (103 pm), potential disorder close to the superconducting plane still introduces significant disorder. On the other hand, compared with Co doping, which also introduces electrons into the system, degree of disorder to the system is weaker, leading to much milder suppression of $T_c$ at the same average valence of Fe.

For undoped $CaKFe_4As_4$, there are clear crossings of $J_c-H$ curves at different temperatures, which can be described as an unexpected peak effect in the $J_c-T$ curve. Such unexpected enhancement of $J_c$ at higher temperatures is rare, and requires some special explanation. Referring to previous study by Ichinose [26], novel planar defects present in $CaKFe_4As_4$ consist of additional $KFe_2As_2$ planes, and they cause an enhancement of the pinning effect. Usually, $KFe_2As_2$ has no superconductivity above ~4 K [24]. However, due to diffusion of holes in $KFe_2As_2$ planar defects into $CaKFe_4As_4$, the hole density in $KFe_2As_2$ planar defects is expected to be lower than that of $KFe_2As_2$ single crystals, leading to higher $T_c$ in the defect region. If this $T_c$ of $KFe_2As_2$ planar defects is between 10 and 15 K, we would expect strong enhancement of $J_c$ above 15 K, since $KFe_2As_2$ planar defects start to work as effective pinning centers above 15 K. In samples with low doping levels, we can still observe crossings of $J_c-H$ curves since the effect of pinning centers caused by doping is not still significant compared with that by planar defects.

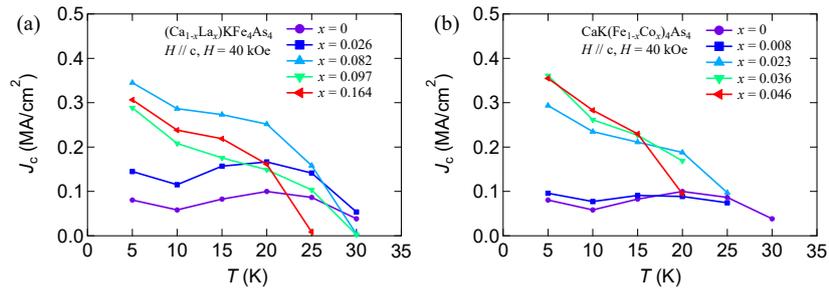

FIG. 8. $J_c$ at 40kOe as functions of temperature in (a) $(Ca_{1-x}La_x)KFe_4As_4$ and (b) $CaK(Fe_{1-x}Co_x)_4As_4$ single crystals with different doping levels $x$.

Now, we compare the impact of two different dopings, La and Co, on the peak effect in the $J_c-T$ curve. As shown in Fig. 8, with increasing doping, $J_c$ at high fields increases significantly in both cases, especially at low temperatures. At low-doping levels, $J_c$ increases with temperature in the range of 10 K < $T$ < 25 K. This anomalous temperature dependence is another manifestation of crossings of $J_c-H$ curves as we have seen in Figs. 4(a)-(e). For Co doping, with increasing doping, the peak effect in the $J_c$–$T$ curve is strongly suppressed, and the field dependence of $J_c$ gradually becomes similar to that in $(Ba_{1-x}K_x)Fe_2As_2$ showing no peak effects[20]. Point defects introduced by Co doping play a dominant role in this case. On the other hand, for La doping, enhancement of $J_c$ at high fields in the range of 10 K < $T$ < 25 K is stronger. As the doping level increases, although the peak in the $J_c$–$T$ curve disappears, the temperature dependence of $J_c$ becomes weaker in the same temperature range. Vortex pinning in this temperature range is controlled by two factors. One is the planar defects that makes $CaKFe_4As_4$ to have a novel peak effect, and the other is the strength of superconductivity induced by doped elements. It is the latter factor that makes the magnitude of peak in $J_c$–$T$ curve stronger. Comparing the two different dopings, their maximum $J_c$ at low temperature is very similar, and we believe that the point defects they introduce have similar effects. Compared with the Co-doped sample, we believe that the suppression of superconductivity brought by La doping is smaller, that is, its enhancement effect on $J_c$ at higher temperatures is more effective. As the temperature increases, some superconducting planar defects become normal. In La-doped samples, superconductivity in the region surrounding the planar defects is stronger than Co-doped sample, resulting in a more effective pinning effect.

## 5. Summary

We have successfully grown single crystals of $(Ca_{1-x}La_x)KFe_4As_4$ ($0 \leq x \leq 0.16$) using self-flux method and investigated the evolution of physical properties including $J_c$. We found that the $c$-axis length does not change significantly with the doping amount. Sharp superconducting transition with flat magnetization at low temperatures even after La-doping demonstrates that the grown crystals are of high quality. By measuring several samples with different doping level, we found that the anisotropy did not change significantly with doping. By measuring the Hall coefficient, we confirm that the electrons are successfully introduced by La-doping. As $x$ increases, $J_c$ at the same temperature and magnetic field increases initially and reaches its maximum at $x = 0.082$, and the anomalous temperature dependence of $J_c$ in $CaKFe_4As_4$ is wiped away. Furthermore, we compare the effects of electron doping with two different substitutions, Co doping and La doping, on $T_c$, $J_c$, and the exponent α related to the magnetic field dependence of $J_c$. By measuring the magnetization for magnetic field parallel to the $ab$-plane, we observe a huge anisotropy between $J_{c2}$ and $J_{c3}$ similar to the case of $CaKFe_4As_4$. STEM observation clearly demonstrates that there are planar defects similar to those in $CaKFe_4As_4$. By comparing with $(Ca_{1-x}La_x)KFe_4As_4$ with $CaK(Fe_{1-x}Co_x)_4As_4$, we find that doping outside FeAs planes can enhance $J_c$ more effectively.